\documentclass[journal]{IEEEtran}

\usepackage{cite}
\usepackage{amsmath,amssymb,amsfonts}
\usepackage{algorithm}
\usepackage{algpseudocode}
\usepackage{graphicx}
\usepackage{textcomp}
\usepackage{xcolor}
\usepackage{enumitem}
\usepackage{lipsum}
\usepackage{titlesec}
\usepackage [english]{babel}
\usepackage [autostyle, english = american]{csquotes}
\usepackage{subcaption}
\usepackage[numbers, compress]{natbib}
\usepackage{array}
\usepackage{tikz}
\usepackage[hidelinks]{hyperref}
\usepackage{adjustbox}
\usepackage{makecell}
\usepackage{mathtools}
\usepackage{tabularx}
\MakeOuterQuote{"}
\IEEEoverridecommandlockouts 
\algdef{SE}[SUBALG]{Indent}{EndIndent}{}{\algorithmicend\ }%
\algtext*{Indent}
\algtext*{EndIndent}

\titlespacing*{\section}{0pt}{1 ex plus .5 ex minus .2ex}{1 ex plus .2ex}
\titlespacing*{\subsection}{0pt}{1 ex plus .5 ex minus .2ex}{1 ex plus .2ex}
\titlespacing*{\subsubsection}{0pt}{1 ex plus .5 ex minus .2ex}{1 ex plus .2ex}

\def\BibTeX{{\rm B\kern-.05em{\sc i\kern-.025em b}\kern-.08em
    T\kern-.1667em\lower.7ex\hbox{E}\kern-.125emX}}

\begin{document}

\title{Multicollinearity-Aware Parameter-Free Strategy for Hyperspectral Band Selection: A Dependence Measures-Based Approach}

\author{Dibyabha Deb,
\thanks{
\indent D. Deb is with the Department of Food, Agricultural, and Biological Engineering, College of Food, Agricultural, and Environmental Sciences, The Ohio State University, USA, and the Department of Electronics and Communication Engineering, Manipal Institute of Technology Bengaluru, Manipal Academy of Higher Education, Manipal, Udupi-576104, India (e-mail:deb.46@osu.edu).}
       Ujjwal Verma,~\IEEEmembership{Senior~Member,~IEEE},
\thanks{
\indent U. Verma is with the Department of Electronics and Communication Engineering, Manipal Institute of Technology, Manipal Academy of Higher Education, Manipal, Udupi-576104, India (e-mail: ujjwal.verma@manipal.edu)(Corresponding Author).}

}

\maketitle

\begin{abstract}
Hyperspectral bands offer rich spectral and spatial information; however, their high dimensionality poses challenges for efficient processing. Band selection (BS) methods aim to extract a smaller subset of bands to reduce spectral redundancy. Existing approaches, such as ranking-based, clustering-based, and iterative methods, often suffer from issues like sensitivity to initialization, parameter tuning, and high computational cost. This work introduces a BS strategy integrating three dependence measures: Average Band Correlation (ABC) and Mutual Information (MI), and Variance Inflation Factor (VIF). ABC quantifies linear correlations between spectral bands, while MI measures uncertainty reduction relative to ground truth labels. To address multicollinearity and reduce the search space, the approach first applies a VIF-based pre-selection of spectral bands.  Subsequently, a clustering algorithm is used to identify the optimal subset of bands based on the ABC and MI values. Unlike previous methods, this approach is completely parameter-free for hyperspectral band selection, eliminating the need for optimal parameter estimation. The proposed method is evaluated on four standard benchmark datasets: WHU-Hi-LongKou, Pavia University, Salinas, and Oil Spill datasets, and is compared to existing state-of-the-art approaches. There is significant overlap between the bands identified by our proposed method and those selected by other methods, indicating that our approach effectively captures the most relevant spectral features. Further, support vector machine (SVM) classification validates that VIF-driven pruning enhances classification by minimizing multicollinearity. Ablation studies confirm that combining ABC with MI yields robust, discriminative band subsets.

\end{abstract}


\section{Introduction}
Hyperspectral image (HSI) provides an exceptional capability to capture detailed spectral and spatial information across many hundreds or thousands of contiguous bands with very narrow and continuous spectral resolution. The extensive spectral information captured by HSI far exceeds that of traditional RGB images or multispectral images. The rich information provided by HSI has enabled a broad spectrum of applications and the list includes, but is not limited to, precision agriculture and water resource management \cite{water1}, military applications \cite{mil}, target detection \cite{tardet} \cite{tardet2}, medical diagnosis \cite{med} \cite{med2}, environmental monitoring \cite{env} \cite{env2}, and mineral exploration \cite{mineral1} \cite{mineral2}. However, the large number of spectral bands introduces challenges associated with high-dimensional data. Specifically, issues such as redundancy and multicollinearity arise due to highly correlated spectral information among adjacent or neighboring bands. Moreover, the "Curse of Dimensionality" or the "Hughes phenomenon" \cite{curse} highlights the detrimental effects of high dimensionality on the performance of classifiers.

To mitigate these challenges, dimensionality reduction (DR) techniques are necessary. Broadly, there are two main types of DR for HSI: feature extraction (FE) and feature selection (FS). Under FE, the original high-dimensional data is transformed into a lower dimensional subspace using linear or non-linear data transformations \cite{pca1} \cite{pca2} \cite{mvpca}. Although effective, these methods result in the loss of original information as the original spectral characteristics are transformed, and hence, loss of interpretability. In contrast, FS (which can also be called Band Selection (BS)) preserves the original spatial and spectral information and selects a subset of the most informative bands from the original set as the representative bands of the entire dataset. BS can be either supervised or unsupervised methods, according to the need for training samples. Supervised methods are based on annotated samples, and unsupervised methods are based on understanding the patterns in the data itself. Due to the lack of annotated samples and the high cost of obtaining labeled data in HSI, unsupervised methods are quite popular.

The BS methods can be subdivided into ranking-based methods \cite{rank1} \cite{rank2} \cite{rank3}, clustering-based methods \cite{cl1} \cite{cl2} \cite{cl3} and iterative or searching-based methods \cite{s1} \cite{s2}. In ranking-based methods, bands are selected based on a predefined criterion. For instance, Chang and Wang \cite{rank1} introduced a constrained energy minimization criterion to rank bands, while Jia et al. \cite{rank2} proposed a ranking-based clustering technique that prioritizes subsets by integrating similarity and discriminative capacity. Further, Xu et al. \cite{rank3} incorporated structural similarity measures for efficient ranking. Clustering-based methods cluster the bands and from each cluster a representative band is selected. In \cite{cl3}, the authors utilized information theoretic metrics to group spectrally similar bands, and Yuan et al. \cite{cl2} used a dual-clustering approach to incorporate contextual spatial information along with spectral information. In searching-based methods, the algorithm iteratively selects bands into the subset based on the optimization of the objective function. Wang et al. \cite{s1} introduced an optimal neighborhood reconstruction approach to preserve local data geometry, while Sui et al. \cite{s2} developed a manifold-preserving method that penalizes redundancy. Although these methods provide high performance, they are often prone to initialization problems which leads to sub-optimal selection of band subset, sensitive to different parameter and hyper-parameter settings and require fine-tuning and are computationally expensive. 

In recent years, to address these gaps, statistical measure-based approaches have emerged as a promising solution. Researchers have worked on the development of BS strategies using statistical measures, such as in \cite{sparse}, where the authors introduced a spectral correlation representation based on sparse reconstruction. A separate study explored the use of correlation coefficients (CC) to effectively select bands in \cite{debcorr}. It introduced a threshold-based selection of bands based on the average correlation values. Building on this work, we introduce a BS strategy based on Average Band Correlation (ABC) and Mutual Information (MI) with pre-selection of a subset of bands using the Variance Inflation Factor (VIF). The proposed ABC-MI method is based on the average correlation values, which capture the linear inter-band relationship, and mutual information, which quantifies the non-linear relationship between the spectral band and the ground truth. In this work, VIF acts as a pre-selection criterion which effectively removes bands with a higher VIF value (based on pairwise values) and reduces the search space for the selection of a final subset of bands using ABC-MI analysis. This achieves two main advantages - (1) reduction in space to perform the ABC-MI analysis and (2) elimination of bands that definitely possess some degree of correlation. Based on the ABC and MI values of the pre-selected bands and treating them as input features, a clustering algorithm is applied. This groups the data points based on ABC and MI values, and, based on the cluster centroids, the closest band to the centroid is selected as its representation.

The major contributions of this paper are as follows.
\begin{enumerate}
    \item The proposed work integrates three statistical dependence measures: ABC, MI, and VIF for band selection. Our experimental results demonstrate the effectiveness of the proposed approach in the identification of key spectral bands across datasets, balancing information richness and redundancy.

    \item VIF-based pre-selection, effectively reduces multicollinearity, enhancing classification performance.
    
    \item Unlike previous methods, this approach is completely parameter-free for hyperspectral band selection and eliminates the need for any optimal parameter estimation.
    
    \item Extensive comparison is conducted with other state-of-the-art methods and on four publicly available datasets that demonstrate the effectiveness of our proposed method.
    
\end{enumerate}
The rest of the article is organized as follows. Section II focuses on the existing state-of-the-art methods and the current literature review. Section III introduces the key theoretical concepts used in our proposed strategy, and Section IV describes the proposed methodology in detail. Sections V and VI explain the experiments and results obtained in detail. In Section VII, the paper is summarized. 
\section{Related Work}

This section briefly discusses recent advances in the field of HSI processing using statistical metrics and information-theoretic measures. A detailed and comprehensive review of different BS strategies is available in \cite{review}.

Early foundational work by Chang et al. \cite{mvpca} introduced a band-prioritization and band-decorrelation approach to band selection based on the spectral decomposition and subsequent ranking using maximum variance principal component analysis (MVPCA). In the information theoretic domain, the authors in \cite{mi1} proposed a ranking-based MI method to select the top bands based on the MI values between the band and the estimated groundtruth map. In the same work \cite{mi1}, they also used two other parameters to control the bands selected in the final subset. These parameters handle the inter-band correlation in a defined bandwidth and complementary information provided by an adjacent band, despite the band falling under the defined bandwidth. Later, Chang et. al \cite{smi} proposed a self mutual information (SMI) based method where the MI was obtained between the band and the entire band set. This method introduced three different variants for the construction of band channel - spectral information divergence (SMI-SID), spectral angle mapper (SMI-SAM), and SIDAM, which is a combination of SID and SAM (SMI-SIDAM).

In \cite{channel}, the authors proposed the band selection problem as a channel capacity problem, treating the entire band set as the input channel and the final subset as the output channel. Sequential (SQ-CCBSS) and successive (SC-CCBSS) were the two variants developed for finding the final subset. However, most of the time, these algorithms run a small number of selected candidates of band subsets due to the computationally exhaustive nature of the possible band subset combinations. In such a scenario, these algorithms lead to a sub-optimal selection of the band subset. Yang et al. \cite{meac} developed a minimum estimated abundance covariance (MEAC) method where the method selects bands which minimizes the covariance between the abundance vector and the estimated abundance vector.

In \cite{fn}, a coarse-fine strategy (FNGBS) is proposed, where a fast neighborhood grouping method extracts context information over a large spectrum range using information entropy. In \cite{ocf}, an optimal clustering framework (OCF) was proposed to search for the optimal clustering structure, and then a ranking-based algorithm was employed to select the representative bands.

The authors in \cite{cl3} proposed a clustering-based hyperspectral band selection method utilizing information measures within an agglomerative hierarchical clustering framework. The core idea was to compute a dissimilarity matrix using either band pairwise normalized MI or symmetric Kullback-Leibler divergence. Hierarchical clustering iteratively merges the most similar band clusters based on complete linkage. Based on the two metrics, the cluster representative was selected based on the highest MI values or the highest divergence values compared to the other bands in the cluster. Xu et al. \cite{rank3} proposed a novel approach for hyperspectral band selection by leveraging structural similarity (SSIM) to measure inter-band relationships and a similarity-based ranking (SR) strategy to evaluate band representativeness. SSIM index was expressed in terms of luminance, contrast, and structure, and evaluated the potential of each band to be a cluster center based on its average similarity and dissimilarity to other bands. All bands were ranked according to the product of the average similarity and dissimilarity, and the final subset of bands was selected based on the top desired number of bands. 

A sparse self-representation (SSR) based method was proposed in \cite{ssr} and in \cite{sparse}, the authors proposed an iterative optimization framework that addresses the limitations of traditional SSR (SCDBS). Based on pairwise Correlation Coefficient, a correlation-derived weight matrix was computed, and the sparsity coefficient matrix was weighted on this. The proposed approach had a higher prioritization for overall high correlation, and to remove redundancy between higher correlation bands, the inter-band correlation was minimized. Cai et al. proposed a graph convolution SSR (GCSR) in \cite{graph} where the relationship between bands and the sparse coefficient matrix was used for clustering. Yuan et al. \cite{mdpp} introduced a multigraph determinantal point process (MDPP) which treats every band as a node and the edge as a relationship between the bands. To capture the intricate nature of the relationship, multiple graphs are designed. A dominant-set-extraction-based selector (DSEBS) was proposed by the authors in \cite{dsebs} which exploits the idea that informative bands must provide well-structured boundaries for different objects in an image. Yu et al. \cite{cscbs} proposed a class signature-constrained background suppression band prioritization (CSCBS-BP). The method uses the idea of linearly constrained minimum variance (LCMV) by interpreting the LCMV as CSCBS by specifying the signal arrival direction. There are two variants - Forward CSCBS-BP (FCSCBS-BP) and Backward CSCBS-BP (BCSCBS-BP). In a similar work, Yu et al. \cite{lcmv} proposed LCMV-based band selection methods where the method interprets the signal sources as class signature vectors and linearly constrains the class signature vectors. There are two variants of this algorithm - LCMV sequential feed forward BS (LCMV-SFBS) and LCMV sequential backward BS (LCMV-SBBS) . 

A band correlation algorithm was proposed in \cite{bca} where the core idea was to iteratively select bands based on a scoring function. The scoring function quantifies the trade-off between the representative ability and redundancy of the band. For each candidate band, the average correlation was calculated between the band and other selected bands and subtracted from the average correlation of the band and other unselected bands. The highest-scoring band was added in the final subset. In \cite{debcorr}, a correlation-based algorithm that calculates the ABC of the band based on the average correlation of the band with respect to all other bands was proposed. Based on the ABC values, a threshold was set, and all the bands below the threshold were selected. Such an approach allowed us to select the bands with less average correlation, indicating the independent nature.

In this paper, we propose the ABC-MI method for hyperspectral band selection with VIF pre-selection. Unlike the approach in \cite{bca}, where correlation is calculated with respect to individual bands, we compute the Average Band Correlation (ABC) considering all bands collectively. Furthermore, unlike \cite{mi1}, which uses an estimated ground-truth map for computing Mutual Information (MI), we utilize the actual ground-truth labels, improving the reliability of MI calculation. Compared to \cite{smi}, which calculates MI between the entire band set, our method focuses on MI between each individual band and the ground truth, enhancing discriminative power.

A key advantage of our method is that it is completely parameter-free, removing the need for fine-tuning or optimal parameter estimation, which is often required in most existing methods. Additionally, we introduce a pre-selection step based on the Variance Inflation Factor (VIF) to effectively control multicollinearity among the selected bands, thereby improving stability and reducing redundancy.

Other state-of-the-art methods like MDPP, DSEBS, SQ-CCBSS, and SC-CCBSS typically operate on a limited number of candidate subsets due to their high computational complexity. This constraint often leads them to suboptimal final band subsets. In contrast, our method scales efficiently by employing VIF-guided pruning and leveraging ABC and MI in a clustering framework, which facilitates the selection of more optimal and representative band subsets.

\section{Background Theory}

The proposed approach employs three statistical dependence metrics to select a subset of bands, aiming to reduce spectral redundancy. These metrics are: Average Band Correlation (ABC), Mutual Information (MI), and Variance Inflation Factor (VIF).

Average Band Correlation (ABC) measures the linear relationship between pairs of spectral bands. A higher ABC value indicates a stronger dependency between the bands, which suggests greater redundancy. On the other hand, Mutual Information (MI) assesses the statistical dependence between the pixel values in a specific band and their corresponding ground truth labels. This metric captures both linear and nonlinear relationships; a higher MI value between a band and its labels signifies greater information content and relevance for classification.

Additionally, the Variance Inflation Factor (VIF) is used to identify multicollinearity among the bands.

Let us assume that an HSI image $\mathcal{I}$ is represented as $\mathcal{I} \in \mathbb{R}^{h \times w \times n}$ where $h \times w$ pixels refer to the spatial dimension and $n$ refer to the number of spectral bands or spectral dimension of the image $\mathcal{I}$, respectively. The set of $n$ bands can be mathematically represented as $\mathbf{B} = [\mathbf{b}_1, \mathbf{b}_2, ..., \mathbf{b}_{n-1}, \mathbf{b}_n]$ and $\mathbf{B} \in \mathbb{R}^{p \times n}$ where $p = h \times w$. To understand the proposed algorithm, we will now discuss the key theoretical concepts used in our strategy.

\subsection{Average Band Correlation}
The correlation coefficient (CC) metric measures the strength and direction of the relationship between any two variables $X$ and $Y$. CC understands the linear relationship between these two variables and quantifies the relationship by ranging the value between -1 to 1, where -1 indicates a perfect negative correlation (if $X$ increases, then $Y$ decreases and vice-versa), 1 indicates a perfect positive correlation (if $X$ increases, then $Y$ increases and vice-versa) and 0 indicates no correlation (if $X$ increases/decreases it does not influence $Y$).

Mathematically, pairwise CC between two variable $X$ and $Y$-can be formulated as 

\begin{equation}
    r_{\mathbf{X,Y}} =\frac{\sum_{i = 1}^{n}{(\mathbf{X} - \bar{\mathbf{X}})(\mathbf{Y} - \bar{\mathbf{Y}})}}{\sqrt{{\sum_{i=1}^{n}{(\mathbf{X} - \bar{\mathbf{X}})^2}}{\sum_{i=1}^{n}{(\mathbf{Y} - \bar{\mathbf{Y}})}^2}}}
     \label{corr}
\end{equation}

In our study, we have defined the two variables, $X$ and $Y$, as the $i$-th and $j$-th band $(\mathbf{b_i, b_j})$ where $i,j \in [1, 2,...,n]$. The CC analysis gives us a $n \times n$ matrix which contains the pairwise CC values between each band of the image $\mathcal{I}$.

For our proposed algorithm, we defined a metric called Average Band Correlation (ABC), and the ABC of $i$-th band can be defined as the mean of the absolute correlation values between band $i$ and band $j$ where $i,j \in [1, 2, ..., n]$ and $j \neq i$. Mathematically, 
\begin{equation}
    ABC_{\mathbf{b_i}} = \frac{1}{n-1}\sum_{j=1,j\neq i}^{n} |r_{\mathbf{b_i},\mathbf{b_j}}|
\label{abc}
\end{equation}
where $|\cdot|$ denotes the absolute value. The ABC values give us a $n \times 1$ vector. The higher the ABC value for a particular band, the more linear dependency the band has with other bands. This means that the band is a good representation of the overall dataset, but the selection of higher ABC valued bands can lead to a less diverse subset. Also, higher ABC valued bands are redundant with other higher ABC valued bands, and bands with lesser ABC values may be a better representation of the dataset.

\subsection{Mutual Information}
From information theory, a random variable $X$ which has a probability distribution set $\mathcal{X} = [\mathcal{X}_1, ..., \mathcal{X}_N]$ will have an entropy which is defined by
\begin{equation}
    H(X) = -\sum_{i=1}^{N} \mathcal{X}_i \log_2\mathcal{X}_i
\label{ent}
\end{equation}
Entropy explains the uncertainty associated with the random variable. Based on this, if for two random variables, $X$ and $Y$, the probability distribution sets are defined as such, $\mathcal{X} = [\mathcal{X}_1, ..., \mathcal{X}_N]$ and $\mathcal{Y} = [\mathcal{Y}_1, ..., \mathcal{Y}_N]$, respectively and the joint probability distribution set is defined as $\mathcal{X, Y} = [\mathcal{(X,Y)}_1, ..., \mathcal{(X,Y)}_N]$ then the Mutual Information (MI) between $X$ and $Y$ can be defined as
\begin{equation}
    I(X, Y) = \sum_{i=1}^{N} \sum_{j=1}^{N} \mathcal{(X,Y)}_{i,j} \log_2\frac{\mathcal{(X,Y)}_{i,j}}{\mathcal{X}_i \mathcal{Y}_j}
\label{mi}
\end{equation}
From the Equations \ref{ent} and \ref{mi}, we can rewrite Equation \ref{mi} in terms of entropy as
\begin{equation}
    I(X,Y) = H(X) + H(Y) - H(X,Y)
\label{mi:ent}
\end{equation}
where $H(X,Y)$ gives the joint entropy of two variables. In our proposed work, the two variables are $i$-th band and label $T$. The motivation behind the use of Mutual Information (MI) was to develop a metric to assess the reduction in uncertainty about the $i$-th band \textit{after} observing label $T$.  This reduction serves as an indicator of the degree of dependency between the $i$-th band and the label $T$. Notably, such a dependency need not be strictly linear and may instead exhibit complex, non-linear characteristics. This non-linear relationship between the $i$-th band of image $\mathcal{I}$ and the corresponding label $T$ can be expressed by the equation 
\begin{equation}
    I(\mathbf{b}_i,T) = H(\mathbf{b}_i) + H(T) - H(\mathbf{b}_i,T)
\label{mi:use}
\end{equation}

Higher mutual information $I(\mathbf{b}_i,T)$ values indicate a stronger statistical dependency between a spectral band $\mathbf{b}_i$ and its ground truth $T$, reflecting the extent to which knowledge of the ground truth reduces the uncertainty (entropy) of the band.

\subsection{Variance Inflation Factor}
The Variance Inflation Factor (VIF) is a statistical measure used in regression analysis to identify multicollinearity, which occurs when independent variables are highly correlated. VIF measures how much the variance of a regression coefficient is inflated due to this correlation. A higher VIF indicates that the predictor’s effect is less reliably estimated because it shares information with other predictors.

In this work, we have used VIF for the pairwise collinearity test to establish a linear relationship between $i$-th and $j$-th bands of the image $\mathcal{I}$. The equation for calculating pairwise VIF between $\mathbf{b_i, b_j}$ can be formulated as
\begin{equation}
    VIF(\mathbf{b}_i, \mathbf{b}_j) = \frac{1}{1-R^2}
\label{vif}
\end{equation}
where $R^2$ is the coefficient of determination. 

Let us assume a random independent variable $X$ and a dependent variable $Y$, each with $m$ observations. The core idea behind linear regression analysis tells us that we can represent $Y$ in terms of $X$, such as 
\begin{equation}
    Y_i = \alpha + \beta X_i + \epsilon_i, \forall i\in m
\label{lr}
\end{equation}
where $\alpha$ is the y-intercept, $\beta$ is the slope of the line, and $\epsilon_i$ is the error between the estimate and the actual observation for the $i$-th observation. Let the ordinary least squares (OLS) estimator for $\alpha$ and $\beta$ be $\hat{\alpha}$ and $\hat{\beta}$, respectively. Then we can write the $Y$ estimate as $\hat{Y}$ and define it as
\begin{equation}
    \hat{Y}_i = \hat{\alpha} + \hat{\beta}  X_i, \forall i \in m
\label{lr-new}
\end{equation}
The coefficient of determination can be, then, defined as
\begin{equation}
    R^2 = \frac{\sum_{i=1}^m (\hat{Y_i} - \bar{Y})^2}{\sum_{i=1}^m (Y_i - \bar{Y})^2}
\label{coeffdet}
\end{equation}
and using $\hat{\beta} = \frac{\sum_{i=1}^m (X_i - \bar{X})(Y_i-\bar{Y})}{\sum_{i=1}^m (X_i - \bar{X})^2}$ and $\hat{\alpha} = \bar{Y} - \hat{\beta}  \bar{X}$, we can rewrite Equation \ref{coeffdet} by combining it with Equation \ref{lr-new} and Equation \ref{corr}
\begin{equation}
    \begin{aligned}
        R^2 & = \frac{\sum_{i=1}^m (\hat{\alpha} + \hat{\beta}  X_i - \bar{Y})^2}{\sum_{i=1}^m (Y_i - \bar{Y})^2}\\
              & = \frac{\sum_{i=1}^m (\bar{Y} - \hat{\beta}  \bar{X} + \hat{\beta}  X_i - \bar{Y})^2}{\sum_{i=1}^m (Y_i - \bar{Y})^2}\\
              & = \frac{[\sum_{i=1}^m (X_i - \bar{X})(Y_i-\bar{Y})]^2}{\sum_{i=1}^m (X_i - \bar{X})^2\sum_{i=1}^m (Y_i - \bar{Y})^2}\\
              & = (\frac{[\sum_{i=1}^m (X_i - \bar{X})(Y_i-\bar{Y})]}{\sqrt{\sum_{i=1}^m (X_i - \bar{X})^2}\sqrt{\sum_{i=1}^m (Y_i - \bar{Y})^2}})^2 \\
              & = r_{X,Y}^2
    \end{aligned}
\label{coeffdet-new}
\end{equation}
A detailed proof is presented in \cite{proof} to prove that the coefficient of determination, $R^2$, is equal to the square of the correlation coefficient, $r_{X,Y}^2$. To reduce the complexity, the square of the calculated pairwise CC values can be used for the coefficient of determination values. Hence, pairwise VIF between $\mathbf{b_i, b_j}$ can be re-formulated (Equation \ref{vif})  as
\begin{equation}
    VIF(\mathbf{b}_i, \mathbf{b}_j) = \frac{1}{1-r_{\mathbf{b}_i, \mathbf{b}_j}^2}
\label{vif-new}
\end{equation}
The rationale for using VIF is to identify a subset of band pairs that yield the minimum VIF values, and consequently, the lowest pairwise correlation. After removing any duplicates, this refined subset of bands is then used for further analysis in the proposed algorithm.  In the context of selecting bands using pairwise VIF, it may happen that $\mathbf{b}_i$ and $\mathbf{b}_j$ are selected based on their values while, $\mathbf{b}_{i+k}$ and $\mathbf{b}_{j+k}$ are selected based on their relative values. However, it may happen that $\mathbf{b}_i$ and $\mathbf{b}_{i+k}$ result in a higher relative VIF value and thus, introduces collinearity in the subset. To address this issue, the proposed strategy involves the elimination of any band that exceeds the $VIF_{lim}$ value to ensure that a subset of bands is selected with controlled collinearity. It is acknowledged that perfect independence among bands may not be possible and achievable at this stage. Therefore, a certain degree of tolerance towards collinearity becomes necessary. This tolerance facilitates a trade-off balance between minimizing multicollinearity and retaining relevant spectral information by ensuring diversity among the selected bands.

\section{Methodology}
As discussed in Section III, let us assume the set of $n$ bands can be represented as $\mathbf{B} = [\mathbf{b}_1,\ldots,\mathbf{b}_n]$ and $\mathbf{B} \in \mathbb{R}^{p \times n}$ where $p = h \times w$. During the pre-processing stage, the analysis is restricted to the pixels that do not belong to the background class, and we can define the remaining pixels by $p' = h' \times w'$. These pixels, $p'$, are then retained by both the dataset and the ground truth $T$. Following the retention of the pixels, the data undergoes bandwise standardization before any subsequent computation. The proposed BS strategy selects a subset of bands from $\mathbf{B}$ which can be represented as $\mathbf{B'} = [\mathbf{b'}_1,\ldots,\mathbf{b'}_{n'}]$ where $n'$ is the total number of bands selected from the set of bands $\mathbf{B}$ and $1 \leq n' \leq n$. The proposed approach first employs the VIF metric to preselect a subset of bands, thereby reducing multicollinearity. Next, the measures of ABC and MI are computed, and these values are subsequently provided as inputs to the k-means clustering algorithm to identify the final set of bands (Algorithm \ref{alg}).

\begin{algorithm}
    \caption{Proposed Algorithm}
    \begin{algorithmic}[1]
        \Require $n =\text{Number of Bands in the Original Image}$,\\
         $\mathbf{B}= [\mathbf{b}_1,\ldots,\mathbf{b}_n]$, \\ 
        $T =\text{Groundtruth}$,\\
        $y =\text{tolerance factor for VIF}$, \\
        $n' =\text{Number of Bands to be selected}$
        \Ensure $\mathbf{B'}= [\mathbf{b'}_1,\ldots,\mathbf{b'}_{n'}]$
        \State Compute CC (Equation \ref{corr}) and VIF (Equation \ref{vif-new})  $\forall ~\mathbf{b}_{i}, \mathbf{b}_{j} \in \mathbf{B}$
        \State Compute ABC (Equation \ref{abc})  $\forall ~\mathbf{b}_{i} \in \mathbf{B}$
        \State Adjust $VIF_{lim}$ (Equation \ref{viflim})
        \State Initialize: $\mathbf{B''} \gets \emptyset$
        \For{$\forall i,j\in[1,...,n], i \neq j$}
        \If{$VIF(\mathbf{b}_i,\mathbf{b}_j) \leq VIF_{lim}$}
        \If{$\mathbf{b}_i \notin \mathbf{B''}$}
        \State $\mathbf{B''} \gets \mathbf{B''} \cup \mathbf{b}_i$
        \EndIf
        \If{$\mathbf{b}_j \notin \mathbf{B''}$}
        \State $\mathbf{B''} \gets \mathbf{B''} \cup \mathbf{b}_j$
        \EndIf
        \EndIf
        \EndFor
        \State Initialize: $\mathbf{B'} \gets \emptyset$
        \State Compute MI (Equation \ref{mi:use})  $\forall \mathbf{b}_i \in \mathbf{B''}$
        \State Construct $\mathbf{ABC-MI} \gets \{ (\text{ABC}_{\mathbf{b}_i}, \text{MI}_{\mathbf{b}_i} )\mid \mathbf{b}_i \in \mathbf{B''} \}$
        \State Apply K-Means$(\mathbf{ABC-MI}, n')$ $\Rightarrow$ obtain cluster assignments for each point ($c^{(1)}, \ldots, c^{(n'')}$) and  centroids ($\mu_1, \ldots, \mu_{n'}$). 
        \For{$k=1$ to $n'$}
        \State $i_{min} = \underset{i:c^{(i)} = k}{\mathrm{argmin}}\|\mathbf{(ABC-MI)}_i - \mu_k \|^2$
        \State $\mathbf{B'} \gets \mathbf{B'} \cup \mathbf{b}_{i_{min}}$
        \EndFor
    \end{algorithmic}
\label{alg}
\end{algorithm}

\textit{VIF-Based Pre-selection:} To quantify the redundancy present in the dataset among each spectral band and reduce multicollinearity, we employ VIF and compute the pairwise VIF values for each band. To compute the VIF of $i$-th band with $j$-th band, we follow Equation \ref{vif-new}. An important point to be noted here is that between any two independent features (or bands, in our work), the minimum VIF ($VIF_{min}$) value will be 1 and cannot be less than 1. To incorporate a degree of tolerance or relaxation in our work, we added a tolerance factor $y$. The rationale behind adding a factor $y$ is to introduce controlled collinearity in the pre-selection subset. In fact, setting the value of $y$ at 0.00, we achieve the threshold-free, parameter-free variation of our proposed algorithm. However, even by setting the value of $y=0.00$, we can still expect collinearity to "leak" into the pre-selection subset. Thus, $y$ acts as a stop-cock to allow the level of collinearity to flow into the pre-selection subset, as different datasets have different characteristics. As discussed earlier, this tolerance facilitates a trade-off balance between minimizing multicollinearity and retaining relevant spectral information by ensuring diversity among the selected bands. The adjustment in terms of VIF can, therefore, be formulated as
\begin{equation}
    VIF_{lim} = VIF_{min} \times (1+\frac{y}{100}) = (1+\frac{y}{100})
\label{viflim}
\end{equation}
Essentially, the threshold-free, parameter-free variation ($y=0.00$) of our algorithm has $VIF_{lim} = VIF_{min} = 1$. Band pairs whose $VIF$ value is less than or equal to $VIF_{lim}$ are considered non-redundant and are retained in a candidate subset $\mathbf{B''}$. We understand that consecutive bands with high correlation can still become part of the candidate subset $\mathbf{B''}$, but we are not concerned about it at this point in our analysis. Rather, we are trying to achieve the elimination of obvious redundant bands. To identify these bands, they do not have any pairwise VIF value (i.e., with any other band) less than $VIF_{lim}$. The total number of bands pre-selected can be represented by $n''$ and $1 \leq n' \leq n'' \leq n$.

\textit{Computation of ABC and MI:} While VIF filters out redundant bands to an extent, it does not assess the utility of any of the bands in the candidate subset $\mathbf{B''}$. Thus, we compute two additional scores – ABC (Equation \ref{abc}) and MI (Equation \ref{mi:use}). As ABC quantifies the linear relationship between the bands, all the bands are considered for ABC computation and hence, performed before the VIF pre-selection. The MI values are computed for the bands in the candidate subset. Together with ABC and MI, each band, $\mathbf{b}_{i}$, is represented as a two-dimensional (2D) point - (ABC, MI). Here, we will refer to the complete 2D space as ABC-MI space - (ABC$_{\mathbf{b}_{i}}$, MI$_{\mathbf{b}_{i}}$$\mid \mathbf{b}_{i} \in \mathbf{B''}$).

\textit{K-means clustering:} To identify a representative and balanced subset of bands from $\mathbf{B''}$, we apply K-Means clustering to the ABC-MI space. We can define a function that computes the clustering algorithm as K-Means($\mathbf{A}$, $k$) where $\mathbf{A}$ is a matrix of size $M \times N$ where $M$ represents the number of samples and $N$ represents the number of features and $k$ is the number of clusters we want to divide the $M$ points into. Additionally, the clustering algorithm returns the centers of the $k$ clusters, effectively returning the updated cluster centroids. We applied the clustering algorithm to the subset of bands identified after VIF criteria. The bands are represented in terms of ABC and MI, where $M$ becomes the number of bands selected after the VIF criteria, and $N$ becomes 2, ABC, and MI values.

A critical point to implement K-Means clustering is the flexibility it offers in selecting the number of clusters, $k$. In our context, $k$ becomes the number of bands we want to select from the dataset. This adaptability makes K-Means clustering particularly powerful compared to other unsupervised clustering algorithms. The algorithm effectively groups the bands based on similarities in their ABC and MI values. Moreover, we run the clustering algorithm with multiple initializations to obtain the best solution. The K-Means clustering algorithm can be defined by considering we have $M$ points, $[x^{(1)}, \ldots, x^{(M)}]$, and each $x^{(i)}$ has $N$ features presented by $x^{(i)} = (x_{1}^{(i)}, \ldots, x_{N}^{(i)}), \forall i \in M$.

The number of clusters, $k$, is determined by the user and is equal to $n'$. The clustering algorithm, K-Means($\mathbf{ABC-MI},\quad n'$), returns the updated centroids of each cluster based on proximity. The final subset that contains the representative bands from each cluster is the desired result $\mathbf{B'}$.

\section{Experiments}
\subsection{Data Sets}
The proposed algorithm was evaluated on the following datasets: 
\begin{enumerate}[font=\itshape]
    \item \textit{Pavia University: }The Pavia University (PA) scene was collected by the Reflective Optics System Imaging Spectrometer system. It is taken over Pavia in northern Italy and contains 103 bands (after discarding bands with low signal-to-noise ratio) and the size of the dataset becomes 610 $\times$ 340 pixels. In total, the entire dataset has 9 classes of land cover objects.
    
    \item \textit{Salinas: }The Salinas (SA) dataset was recorded by the Airborne Visible / Infrared Imaging Spectrometer over the Salinas Valley in California, United States of America. It contains 204 bands (after discarding bands with low signal-to-noise ratio) and the size of the dataset is 512 $\times$ 217 pixels. There are a total of 16 classes of interest in the image in the dataset. 

    \item \textit{LongKou: }The WHU-Hi-LongKou (LK) dataset was acquired with an 8-mm focal length Headwall Nano-Hyperspec imaging sensor equipped on a DJI Matrice 600 Pro (DJI M600 Pro) UAV platform. The dataset was captured over a small town in LongKou, Hubei province of China. It contains 270 bands and the size of the dataset is 550 $\times$ 400 pixels and has 9 classes of interest \cite{lk1} \cite{lk2}.

    \item \textit{Oil Spill: }The Oil Spill (OS) dataset was captured over the Gulf of Mexico using the Airborne Visible / Infrared Imaging Spectrometer across different test sites \cite{os}. There are only 2 classes present in this dataset - oil and water. For our study, we used GM17 and originally, it had 224 bands. After careful inspection, 34 bands had to be removed due to noisy data, and in total, we had 190 bands. The bands removed were 107-116, 152-170, and 220-224. The size of the dataset is 600 $\times$ 400 pixels. 
    
\end{enumerate}
PA and SA can be accessed from the given link ($\href{https://www.ehu.eus/ccwintco/index.php/Hyperspectral_Remote_Sensing_Scenes}{pa-sa-data}$) \cite{dataset}. LK can be accessed from the link ($\href{https://rsidea.whu.edu.cn/resource_WHUHi_sharing.htm}{lk-data}$) \cite{lkdata} and OS can be obtained from the link ($\href{https://ieee-dataport.org/documents/hyperspectral-remote-sensing-benchmark-database-oil-spill-detection-isolation-forest}{os-data}$) \cite{osdata}. 

\subsection{Classification Setup}
The quality of the bands obtained using our method was assessed by using the support vector machine (SVM) classifier. In these experiments, 10$\%$ of the samples (or pixels) from each class are taken in the training set and the rest in the testing set. Since the distribution of the samples can be random, the final result is an averaged value of ten individual runs to reduce the effect of randomness. The classification performance is assessed by taking the overall accuracy (OA) and Cohen's Kappa score (Kappa) as the chosen metrics for comparison. 
OA can be expressed as 
\begin{equation}
    OA = \frac{\text{total number of pixels correctly classified}}{\text{total number of pixels in the dataset}}
\label{oa}
\end{equation}
Kappa can be expressed as
\begin{equation}
    Kappa = \frac{p_o -p_e}{1-p_e}
\label{kappa}
\end{equation}
where $p_o$ refers to the observed agreement and $p_e$ refers to the hypothetical random agreement. 
We used the radial basis function (RBF) kernel, and the hyperparameters C and gamma are optimized at each run during the training phase by using grid search. 

The experiment is conducted by varying the value of $n'$ from 5 to 50 with a step size of 5. The value of $y$ was varied between 0.00, 0.01, and 0.05 for each value of $n'$ for the PA, SA, and OS datasets, and for LK, it was varied between 0.3, 0.5, and 1. For the K-Means clustering algorithm, it was initialized 40 times with different seeds of centroids at each initialization, returning the best output.

\section{Results and Discussion}
The proposed approach is compared with the following state-of-the-art band selection approaches: MVPCA \cite{mvpca}, FNGBS \cite{fn}, OCF \cite{ocf}, SR \cite{rank3}, SSR \cite{ssr}, GCSR \cite{graph}, SCDBS \cite{sparse}, SMI-BS \cite{smi}, SQ-CCBSS \cite{channel}, MEAC \cite{meac}, MDPP \cite{mdpp}, DSEBS \cite{dsebs}, LCMV \cite{lcmv}, FCSCBS-BP/BCSCBS-BP \cite{cscbs} and UBS \cite{UBSMatlab}.  We first present the bands selected using the proposed approach and other approaches in SA (Table \ref{sabands}), and PA (Table \ref{pabands}). There is significant overlap between the bands selected using the proposed approach (ABC-MI) and those chosen by other methods. For the PA dataset, common bands such as 1, 25, 69, 76, 77, 78, 79, 83, 85, 101, and 103 frequently appear in both ABC-MI and several alternative approaches, especially MEAC, DSEBS, LCMV-SBBS, BCSCBS-BP, and UBS. A similar trend is observed in the SA dataset: the following bands frequently appear in the proposed approach (ABC-MI) and several existing approaches: 3, 96, 193, 107, and 23. The substantial overlap between the bands identified by our proposed method and those selected by existing approaches demonstrates that our method effectively captures the most relevant and informative spectral features. Note that the selected bands are available only for a few studies, as reported in Tables \ref{sabands}, \ref{pabands}.

\begin{table*}[htbp]
    \centering
    \caption{\\COMPARISON OF BANDS SELECTED USING DIFFERENT METHODS FOR SA DATASET. THE PROPOSED APPROACH IS SHOWN AS ABC-MI ($y$).}
    \setlength{\tabcolsep}{2pt}
    \renewcommand{\arraystretch}{3}
    \begin{adjustbox}{width=0.8\textwidth}
    \begin{tabular}{c|ccccccccccccccccccccc|c}
    \hline
        \textbf{METHOD} & \multicolumn{21}{c}{\textbf{BAND NUMBERS (ABC-MI METHODS: 20; REST: 21)}} & \textbf{OA} \\ \hline
        \textbf{ABC-MI (0.00)} & 3 & 4 & 6 & 23 & 37 & 39 & 56 & 68 & 80 & 82 & 83 & 96 & 100 & 107 & 114 & 115 & 125 & 151 & 153 & 193 & - & 92.53 \\ \hline
        \textbf{ABC-MI (0.01)} & 3 & 4 & 6 & 9 & 20 & 23 & 39 & 59 & 79 & 80 & 82 & 83 & 96 & 107 & 114 & 115 & 125 & 151 & 153 & 193 & - & 92.19 \\ \hline
        \textbf{ABC-MI (0.05)} & 3 & 9 & 20 & 23 & 39 & 45 & 68 & 71 & 80 & 82 & 83 & 96 & 106 & 107 & 114 & 115 & 125 & 132 & 153 & 193 & - & 92.74 \\ \hline
        \textbf{SMI-BS (SAM) \cite{smi}} & 15 & 25 & 32 & 50 & 72 & 73 & 76 & 109 & 118 & 120 & 126 & 149 & 156 & 172 & 176 & 184 & 185 & 201 & 203 & 206 & 223 & 93.38 \\ \hline
        \textbf{SMI-BS (SID) \cite{smi}} & 15 & 22 & 32 & 34 & 40 & 41 & 63 & 64 & 71 & 94 & 117 & 118 & 150 & 151 & 157 & 176 & 184 & 185 & 202 & 206 & 223 & 93.85 \\ \hline
        \textbf{SMI-BS (SIDAM) \cite{smi}} & 26 & 27 & 31 & 54 & 61 & 94 & 96 & 108 & 118 & 120 & 139 & 148 & 149 & 156 & 161 & 167 & 174 & 193 & 194 & 195 & 223 & 93.21 \\ \hline
        \textbf{SQ-CCBSS \cite{channel}} & 42 & 43 & 44 & 45 & 46 & 47 & 48 & 49 & 50 & 51 & 52 & 178 & 179 & 180 & 181 & 182 & 183 & 184 & 185 & 186 & 206 & 90.59 \\ \hline
        \textbf{SC-CCBSS \cite{channel}} & 116 & 117 & 118 & 119 & 120 & 121 & 122 & 123 & 124 & 125 & 128 & 155 & 156 & 191 & 192 & 193 & 194 & 195 & 196 & 197 & 198 & 92.68 \\ \hline
        \textbf{MEAC \cite{meac}} & 3 & 5 & 8 & 10 & 12 & 17 & 18 & 25 & 28 & 32 & 36 & 44 & 51 & 58 & 68 & 105 & 107 & 110 & 148 & 149 & 203 & 94.22 \\ \hline
        \textbf{MDPP \cite{mdpp}} & 1 & 8 & 11 & 22 & 27 & 28 & 50 & 57 & 58 & 65 & 90 & 99 & 105 & 119 & 123 & 134 & 142 & 157 & 175 & 191 & 204 & 93.65 \\ \hline
        \textbf{DSEBS \cite{dsebs}} & 16 & 17 & 42 & 44 & 46 & 47 & 99 & 101 & 102 & 112 & 119 & 120 & 121 & 131 & 135 & 174 & 175 & 177 & 180 & 187 & 196 & 94.43 \\ \hline
        \textbf{LCMV-SFBS \cite{lcmv}} & 3 & 14 & 28 & 32 & 38 & 44 & 46 & 47 & 56 & 64 & 96 & 107 & 108 & 109 & 155 & 156 & 157 & 158 & 159 & 160 & 163 & 90.91 \\ \hline
        \textbf{LCMV-SBBS \cite{lcmv}} & 2 & 3 & 9 & 10 & 14 & 16 & 17 & 18 & 19 & 20 & 21 & 22 & 23 & 24 & 25 & 26 & 38 & 39 & 40 & 41 & 42 & 92.78 \\ \hline
        \textbf{FCSCBS-BP \cite{cscbs}} & 41 & 42 & 43 & 44 & 45 & 46 & 47 & 48 & 49 & 50 & 51 & 52 & 53 & 54 & 55 & 56 & 57 & 58 & 59 & 60 & 61 & 91.84 \\ \hline
        \textbf{BCSCBS-BP \cite{cscbs}} & 10 & 12 & 32 & 38 & 44 & 84 & 105 & 106 & 117 & 120 & 121 & 156 & 173 & 174 & 175 & 176 & 192 & 195 & 196 & 197 & 220 & 93.98 \\ \hline
        \textbf{UBS \cite{UBSMatlab}} & 1 & 12 & 23 & 34 & 45 & 56 & 67 & 78 & 89 & 100 & 111 & 122 & 133 & 144 & 155 & 166 & 177 & 188 & 199 & 210 & 224 & 93.43 \\ \hline
    \end{tabular}
    \end{adjustbox}
    \label{sabands}
\end{table*}

\begin{table*}[htbp]
    \centering
    \caption{\\COMPARISON OF BANDS SELECTED USING DIFFERENT METHODS FOR PA DATASET. THE PROPOSED APPROACH IS SHOWN AS ABC-MI ($y$).}
    \renewcommand{\arraystretch}{2}
    \begin{adjustbox}{width=0.8\textwidth}
    \begin{tabular}{c|ccccccccccccccc|c}
    \hline
        \textbf{METHOD} & \multicolumn{15}{c}{\textbf{BAND NUMBERS (ABC-MI METHODS: 15; REST: 14)}} & \textbf{OA} \\ \hline
        \textbf{ABC-MI (0.00)} & 1 & 7 & 25 & 28 & 32 & 44 & 60 & 69 & 76 & 77 & 78 & 79 & 83 & 85 & 101 & 93.00 \\ \hline
        \textbf{ABC-MI (0.01)} & 1 & 6 & 8 & 10 & 25 & 45 & 69 & 76 & 77 & 78 & 79 & 83 & 85 & 101 & 103 & 92.24 \\ \hline
        \textbf{ABC-MI (0.05)} & 1 & 2 & 6 & 8 & 10 & 25 & 27 & 46 & 68 & 69 & 76 & 78 & 80 & 85 & 103 & 90.69 \\ \hline
        \textbf{SMI-BS (SAM) \cite{smi}} & 9 & 18 & 21 & 22 & 37 & 46 & 48 & 50 & 57 & 66 & 82 & 91 & 92 & 94 & - & 92.83 \\ \hline
        \textbf{SMI-BS (SID) \cite{smi}} & 9 & 18 & 21 & 22 & 37 & 40 & 41 & 48 & 57 & 66 & 82 & 91 & 92 & 94 & - & 92.79 \\ \hline
        \textbf{SMI-BS (SIDAM) \cite{smi}} & 8 & 16 & 21 & 22 & 26 & 39 & 40 & 43 & 51 & 59 & 74 & 80 & 84 & 91 & - & 91.29 \\ \hline
        \textbf{SQ-CCBSS \cite{channel}} & 16 & 17 & 18 & 19 & 20 & 79 & 80 & 81 & 82 & 83 & 84 & 87 & 88 & 89 & - & 88.98 \\ \hline
        \textbf{SC-CCBSS \cite{channel}} & 54 & 55 & 56 & 57 & 58 & 87 & 88 & 89 & 90 & 91 & 92 & 93 & 94 & 95 & - & 80.68 \\ \hline
        \textbf{MEAC \cite{meac}} & 1 & 23 & 24 & 25 & 31 & 40 & 42 & 47 & 48 & 54 & 56 & 58 & 59 & 83 & - & 85.66 \\ \hline
        \textbf{MDPP \cite{mdpp}} & 2 & 23 & 44 & 46 & 50 & 62 & 66 & 73 & 89 & 91 & 92 & 93 & 96 & 102 & - & 91.89 \\ \hline
        \textbf{DSEBS \cite{dsebs}} & 1 & 6 & 7 & 19 & 20 & 22 & 63 & 64 & 65 & 66 & 67 & 86 & 95 & 102 & - & 91.97 \\ \hline
        \textbf{LCMV-SFBS \cite{lcmv}} & 1 & 4 & 16 & 36 & 43 & 55 & 56 & 57 & 62 & 63 & 74 & 86 & 91 & 92 & - & 91.59 \\ \hline
        \textbf{LCMV-SBBS \cite{lcmv}} & 1 & 2 & 3 & 6 & 11 & 12 & 40 & 72 & 73 & 82 & 83 & 85 & 86 & 87 & - & 93.02 \\ \hline
        \textbf{FCSCBS-BP \cite{cscbs}} & 9 & 10 & 11 & 12 & 13 & 14 & 15 & 16 & 17 & 18 & 19 & 20 & 21 & 22 & - & 69.77 \\ \hline
        \textbf{BCSCBS-BP \cite{cscbs}} & 1 & 66 & 68 & 69 & 76 & 77 & 78 & 79 & 81 & 82 & 93 & 94 & 97 & 101 & - & 92.99 \\ \hline
        \textbf{UBS \cite{UBSMatlab}} & 1 & 9 & 17 & 25 & 33 & 41 & 49 & 57 & 65 & 73 & 81 & 89 & 97 & 103 & - & 94.64 \\ \hline
    \end{tabular}
    \end{adjustbox}
    \label{pabands}
\end{table*}

\textit{Ablation study: }The role of VIF in reducing multicollinearity and its influence on the pre-selection of bands in the proposed approach are summarized in Tables \ref{reduce1} and \ref{reduce2}. These tables report the total number of bands before and after applying VIF-based pre-selection across the four datasets. The results demonstrate that the VIF-based approach substantially reduces the number of spectral bands, thereby reducing multicollinearity. Specifically, VIF-based pre-selection achieves a reduction of approximately 70\% to 30\% of the total bands, depending on the chosen tolerance factor $y$. This tolerance factor, therefore, serves as an effective mechanism for controlling collinearity among the selected bands. As discussed earlier, it determines the permissible level of collinearity within the pre-selected subset, thereby facilitating a trade-off between minimizing multicollinearity and preserving relevant spectral information by ensuring diversity among the selected bands.

\begin{table*}[htbp]
    \centering
    \caption{\\REDUCTION IN BAND SUBSET AFTER VIF PRE-SELECTION FOR PA, SA, AND OS. THE PROPOSED APPROACH IS SHOWN AS ABC-MI ($y$).}
    \renewcommand{\arraystretch}{5}
    \setlength{\tabcolsep}{3pt}
    \begin{adjustbox}{width=\textwidth}
    \begin{tabular}{c|c|c|c|c|c|c|c|c|c}
    \hline
        \shortstack{{}\\ \textbf{METHOD} \\{}} & \shortstack{{} \\ \textbf{TOTAL NO. OF} \\ \textbf{BANDS BEFORE} \\ \textbf{VIF FOR PA}} & \shortstack{{} \\ \textbf{TOTAL NO. OF} \\ \textbf{BANDS AFTER} \\ \textbf{VIF FOR PA}} & \shortstack{{} \\ \textbf{REDUCTION IN} \\ \textbf{BAND SET SIZE} \\  \textbf{(IN \%) FOR PA}} & \shortstack{{} \\ \textbf{TOTAL NO. OF} \\ \textbf{BANDS BEFORE} \\ \textbf{VIF FOR SA}} & \shortstack{{} \\ \textbf{TOTAL NO. OF} \\ \textbf{BANDS AFTER} \\ \textbf{VIF FOR SA}}  & \shortstack{{} \\ \textbf{REDUCTION IN} \\ \textbf{BAND SET SIZE} \\  \textbf{(IN \%) FOR SA}} & \shortstack{{} \\ \textbf{TOTAL NO. OF} \\ \textbf{BANDS BEFORE} \\ \textbf{VIF FOR OS}} & \shortstack{{} \\ \textbf{TOTAL NO. OF} \\ \textbf{BANDS AFTER} \\ \textbf{VIF FOR OS}}  & \shortstack{{} \\ \textbf{REDUCTION IN} \\ \textbf{BAND SET SIZE} \\  \textbf{(IN \%) FOR OS}} \\ \hline
        \textbf{ABC - MI (0.00)} & 103 & 34 & 66.99 & 204 & 72 & 64.71 & 190 & 58 & 69.47 \\ \hline
        \textbf{ABC - MI (0.01)} & 103 & 49 & 52.43 & 204 & 94 & 53.92 & 190 & 87 & 54.21 \\ \hline
        \textbf{ABC - MI (0.05)} & 103 & 77 & 25.24 & 204 & 121 & 40.69 & 190 & 127 & 33.16 \\ \hline
    \end{tabular}
    \end{adjustbox}
    \label{reduce1}
\end{table*}

\begin{table*}[htbp]
    \centering
    \caption{\\REDUCTION IN BAND SUBSET AFTER VIF PRE-SELECTION FOR LK. THE PROPOSED APPROACH IS SHOWN AS ABC-MI ($y$).}
    \renewcommand{\arraystretch}{2}
    \begin{tabular}{c|c|c|c}
    \hline
        \shortstack{{}\\ \textbf{METHOD} \\{}} & \shortstack{{} \\ \textbf{TOTAL NO. OF} \\ \textbf{BANDS BEFORE} \\ \textbf{VIF FOR LK}} & \shortstack{{} \\ \textbf{TOTAL NO. OF} \\ \textbf{BANDS AFTER} \\ \textbf{VIF FOR LK}} & \shortstack{{} \\ \textbf{REDUCTION IN} \\ \textbf{BAND SET SIZE} \\  \textbf{(IN \%) FOR LK}} \\ \hline
        \textbf{ABC - MI (0.3)} & 270 & 50 & 81.48 \\ \hline
        \textbf{ABC - MI (0.5)} & 270 & 90 & 66.67 \\ \hline
        \textbf{ABC - MI (1)} & 270 & 146 & 45.93 \\ \hline
    \end{tabular}
    \label{reduce2}
\end{table*}

The significance of three statistical metrics—ABC, MI, and VIF—in band selection is highlighted in Tables \ref{paab}, \ref{saab}, \ref{lkab}, and \ref{osab}. These tables show the number of bands selected and overall accuracy when using each of these metrics individually, as well as in combination, across the four datasets. For the PA dataset (Table \ref{paab}), the combination of ABC-MI and VIF (0.00) achieves the highest accuracy. Besides, accuracy decreases as the VIF threshold in ABC-MI increases (from 0.00 to 0.05), indicating that more aggressive band reduction may improve classification.  Moreover, methods employing only MI or ABC without VIF have slightly lower accuracies, suggesting that combining MI + ABC + VIF yields better band subsets. A similar trend is observed in the SA dataset (Table \ref{saab}): The accuracy increases when a combination of all three metrics is utilized. However, a competitive performance is observed when only ABC is used as compared to ABC-MI (0.00). For the LK dataset (Table \ref{lkab}), the highest accuracy is achieved by using only ABC (without VIF pre-selection)(97.92\%), closely followed by ABC-MI (1) (97.27\%) and ABC-MI without VIF (97.91\%), indicating the strong performance of ABC-based and combined methods. These results show that combining ABC and MI with less aggressive redundancy removal (or without VIF) seems to balance band informativeness and diversity to maximize classification accuracy. The variation in overall accuracy across the dataset for different values of the VIF tolerance factor $y$ further emphasizes its importance in controlling the collinearity present in the dataset. The ablation study on the OS dataset (Table \ref{osab}) shows all tested band selection methods reliably select discriminative spectral bands, yielding very high classification accuracy. ABC-MI variants with different thresholds (0.00, 0.01, 0.05) have almost indistinguishable accuracies, suggesting parameter choice has minimal impact here. These results demonstrate the effectiveness of the proposed approach in the identification of key spectral bands across datasets, balancing information richness and redundancy. VIF-based pruning effectively reduces multicollinearity, enhancing classification performance, while ablation studies show that combining ABC with Mutual Information yields robust and discriminative band subsets.

\begin{table*}[htbp]
    \centering
    \caption{\\ABLATION STUDY: BANDS SELECTED FOR PA DATASET. THE PROPOSED APPROACH IS SHOWN AS ABC-MI ($y$). THE APPROACH WITHOUT VIF PRESELECTION IS SHOWN AS W/O VIF. }
    \renewcommand{\arraystretch}{3}
    \begin{adjustbox}{width=0.8\textwidth}
    \begin{tabular}{c|cccccccccccccccccccc|c}
    \hline
        \textbf{METHOD} & \multicolumn{20}{c}{\textbf{BANDS SELECTED; 20 BANDS}} & \textbf{OA} \\ \hline
        \textbf{ABC-MI (0.00)} & 1 & 7 & 8 & 25 & 28 & 32 & 34 & 37 & 44 & 60 & 69 & 76 & 77 & 78 & 79 & 80 & 83 & 85 & 98 & 101 & 93.42 \\ \hline
        \textbf{ABC-MI (0.01)} & 1 & 6 & 7 & 9 & 25 & 29 & 32 & 42 & 45 & 60 & 69 & 76 & 77 & 78 & 79 & 80 & 83 & 85 & 101 & 103 & 93.25 \\ \hline
        \textbf{ABC-MI (0.05)} & 1 & 2 & 6 & 8 & 10 & 14 & 25 & 28 & 42 & 46 & 68 & 69 & 76 & 77 & 78 & 79 & 81 & 85 & 98 & 103 & 92.05 \\ \hline
        \textbf{ONLY MI W/O VIF} & 1 & 2 & 3 & 6 & 7 & 9 & 17 & 28 & 32 & 42 & 64 & 69 & 71 & 72 & 75 & 76 & 85 & 93 & 101 & 103 & 92.35 \\ \hline
        \textbf{ONLY ABC W/O VIF} & 1 & 23 & 28 & 31 & 42 & 47 & 51 & 55 & 68 & 69 & 72 & 73 & 74 & 77 & 78 & 79 & 80 & 81 & 83 & 87 & 91.94 \\ \hline
        \textbf{ABC-MI W/O VIF} & 1 & 2 & 6 & 10 & 14 & 44 & 50 & 65 & 68 & 69 & 71 & 73 & 74 & 75 & 76 & 78 & 80 & 85 & 98 & 103 & 92.29 \\ \hline
    \end{tabular}
    \end{adjustbox}
    \label{paab}
\end{table*}

\begin{table*}[htbp]
    \centering
    \caption{\\ABLATION STUDY: BANDS SELECTED FOR SA DATASET. THE PROPOSED APPROACH IS SHOWN AS ABC-MI ($y$). THE APPROACH WITHOUT VIF PRESELECTION IS SHOWN AS W/O VIF.}
    \renewcommand{\arraystretch}{3}
    \begin{adjustbox}{width=0.8\textwidth}
    \begin{tabular}{c|cccccccccccccccccccc|c}
    \hline
        \textbf{METHOD} & \multicolumn{20}{c}{\textbf{BANDS SELECTED; 20 BANDS}} & \textbf{OA} \\ \hline
        \textbf{ABC-MI (0.00)} & 3 & 4 & 6 & 23 & 37 & 39 & 56 & 68 & 80 & 82 & 83 & 96 & 100 & 107 & 114 & 115 & 125 & 151 & 153 & 193 & 92.53 \\ \hline
        \textbf{ABC-MI (0.01)} & 3 & 4 & 6 & 9 & 20 & 23 & 39 & 59 & 79 & 80 & 82 & 83 & 96 & 107 & 114 & 115 & 125 & 151 & 153 & 193 & 92.19 \\ \hline
        \textbf{ABC-MI (0.05)} & 3 & 9 & 20 & 23 & 39 & 45 & 68 & 71 & 80 & 82 & 83 & 96 & 106 & 107 & 114 & 115 & 125 & 132 & 153 & 193 & 92.74 \\ \hline
        \textbf{ONLY MI W/O VIF} & 1 & 2 & 3 & 39 & 57 & 80 & 81 & 98 & 101 & 107 & 113 & 124 & 134 & 150 & 153 & 186 & 193 & 199 & 222 & 223 & 90.65 \\ \hline
        \textbf{ONLY ABC W/O VIF} & 4 & 12 & 22 & 26 & 30 & 37 & 38 & 46 & 60 & 66 & 72 & 84 & 101 & 103 & 104 & 107 & 115 & 125 & 157 & 178 & 92.82 \\ \hline
        \textbf{ABC-MI W/O VIF} & 1 & 3 & 20 & 23 & 24 & 42 & 78 & 80 & 82 & 96 & 103 & 107 & 132 & 153 & 194 & 200 & 220 & 221 & 222 & 223 & 91.24 \\ \hline
    \end{tabular}
    \end{adjustbox}
    \label{saab}
\end{table*}

\begin{table*}[htbp]
    \centering
    \caption{\\ABLATION STUDY: BANDS SELECTED FOR LK DATASET. THE PROPOSED APPROACH IS SHOWN AS ABC-MI ($y$). THE APPROACH WITHOUT VIF PRESELECTION IS SHOWN AS W/O VIF.}
    \renewcommand{\arraystretch}{3}
    \begin{adjustbox}{width=0.8\textwidth}
    \begin{tabular}{c|cccccccccccccccccccc|c}
    \hline
        \textbf{METHOD} & \multicolumn{20}{c}{\textbf{BANDS SELECTED; 20 BANDS}} & \textbf{OA} \\ \hline
        \textbf{ABC-MI (0.3)} & 112 & 114 & 115 & 116 & 117 & 118 & 119 & 159 & 165 & 166 & 167 & 168 & 179 & 181 & 187 & 191 & 194 & 197 & 199 & 201 & 93.91 \\ \hline
        \textbf{ABC-MI (0.5)} & 3 & 49 & 50 & 51 & 112 & 113 & 115 & 118 & 120 & 157 & 159 & 167 & 168 & 169 & 191 & 193 & 199 & 207 & 221 & 223 & 96.77 \\ \hline
        \textbf{ABC-MI (1)} & 1 & 3 & 7 & 8 & 33 & 43 & 51 & 52 & 53 & 107 & 110 & 112 & 113 & 115 & 154 & 167 & 191 & 199 & 239 & 246 & 97.27 \\ \hline
        \textbf{ONLY MI W/O VIF} & 9 & 13 & 15 & 16 & 17 & 19 & 30 & 37 & 52 & 69 & 78 & 91 & 96 & 107 & 123 & 134 & 136 & 195 & 229 & 239 & 97.21 \\ \hline
        \textbf{ONLY ABC W/O VIF} & 1 & 3 & 26 & 31 & 37 & 55 & 58 & 59 & 61 & 65 & 118 & 134 & 138 & 139 & 141 & 144 & 149 & 150 & 161 & 167 & 97.92 \\ \hline
        \textbf{ABC-MI W/O VIF} & 1 & 3 & 8 & 13 & 48 & 51 & 52 & 53 & 57 & 60 & 91 & 99 & 110 & 138 & 141 & 145 & 148 & 156 & 167 & 197 & 97.91 \\ \hline
    \end{tabular}
    \end{adjustbox}
    \label{lkab}
\end{table*}

\begin{table*}[htbp]
    \centering
    \caption{\\ABLATION STUDY: BANDS SELECTED FOR OS DATASET. THE PROPOSED APPROACH IS SHOWN AS ABC-MI ($y$). THE APPROACH WITHOUT VIF PRESELECTION IS SHOWN AS W/O VIF.}
    \renewcommand{\arraystretch}{3}
    \begin{adjustbox}{width=0.8\textwidth}
    \begin{tabular}{c|cccccccccccccccccccc|c}
    \hline
        \textbf{METHOD} & \multicolumn{20}{c}{\textbf{BANDS SELECTED; 20 BANDS}} & \textbf{OA} \\ \hline
        \textbf{ABC-MI (0.00)} & 1 & 2 & 3 & 8 & 14 & 15 & 30 & 39 & 79 & 81 & 104 & 105 & 121 & 153 & 159 & 184 & 213 & 216 & 217 & 219 & 97.67 \\ \hline
        \textbf{ABC-MI (0.01)} & 1 & 2 & 3 & 7 & 14 & 30 & 39 & 81 & 85 & 104 & 105 & 119 & 123 & 153 & 154 & 158 & 159 & 213 & 217 & 219 & 97.66 \\ \hline
        \textbf{ABC-MI (0.05)} & 1 & 2 & 3 & 7 & 14 & 29 & 39 & 81 & 86 & 101 & 104 & 105 & 119 & 142 & 152 & 153 & 154 & 204 & 213 & 219 & 97.65 \\ \hline
        \textbf{ONLY MI W/O VIF} & 1 & 6 & 16 & 20 & 21 & 24 & 34 & 47 & 66 & 78 & 81 & 104 & 105 & 117 & 141 & 153 & 188 & 207 & 213 & 215 & 97.74 \\ \hline
        \textbf{ONLY ABC W/O VIF} & 1 & 6 & 8 & 13 & 16 & 18 & 19 & 20 & 22 & 30 & 41 & 61 & 72 & 84 & 157 & 158 & 184 & 213 & 215 & 219 & 97.68 \\ \hline
        \textbf{ABC-MI W/O VIF} & 1 & 3 & 14 & 16 & 17 & 20 & 23 & 29 & 40 & 68 & 86 & 101 & 104 & 105 & 118 & 152 & 153 & 154 & 213 & 219 & 97.63 \\ \hline
    \end{tabular}
    \end{adjustbox}
    \label{osab}
\end{table*}

\begin{table*}[htbp]
    \centering
    \captionsetup{justification = centering, singlelinecheck = false}
    \caption{\\AVERAGE OVERALL ACCURACY VALUES OVER DIFFERENT BAND NUMBERS (5:5:50) USING SVM. THE PROPOSED APPROACH IS SHOWN AS ABC-MI ($y$). }
    \setlength{\tabcolsep}{5pt}
    \renewcommand{\arraystretch}{4}
    \begin{adjustbox}{width=\textwidth}
    \begin{tabular}{c|c|c|c|c|c|c|c|c|c|c|c}
    \hline
        \textbf{DATASET} & \makecell{\textbf{ALL} \\ \textbf{BANDS}} & \textbf{MVPCA \cite{mvpca}} & \textbf{FNGBS \cite{fn}} & \textbf{OCF \cite{ocf}} & \textbf{SR \cite{rank3}} & \textbf{SSR \cite{ssr}} & \textbf{GCSR \cite{graph}} & \textbf{SCDBS \cite{sparse}} & \makecell{\textbf{ABC-MI} \\ \textbf{(0.00)}} & \makecell{\textbf{ABC-MI} \\ \textbf{(0.01)}} & \makecell{\textbf{ABC-MI} \\ \textbf{(0.05)}} \\ \hline
        \textbf{SA} & 93.18±0.00 & 87.19±4.37 & 92.86±1.40 & 93.02±0.79 & 91.73±1.88 & 90.14±0.67 & 92.93±0.95 & 93.15±0.99 & 91.85±2.45 & 91.91±2.51 & 92.23±1.76 \\ \hline
        \makecell{\textbf{PA}} & 94.41±0.00 & 85.34±9.84 & 91.02±4.72 & 91.42±3.45 & 89.33±4.52 & 90.67±3.75 & 91.63±3.66 & 92.62±2.49 & 91.52±4.02 & 92.38±3.36 & 91.55±3.42 \\ \hline
    \end{tabular}
    \end{adjustbox}
    \label{tab1}
\end{table*}

\begin{table*}[htbp]
    \centering
    \captionsetup{justification = centering, singlelinecheck = false}
    \caption{\\AVERAGE KAPPA VALUES OVER DIFFERENT BAND NUMBERS (5:5:50) USING SVM. THE PROPOSED APPROACH IS SHOWN AS ABC-MI ($y$).}
    \setlength{\tabcolsep}{3pt}
    \renewcommand{\arraystretch}{4}
    \begin{adjustbox}{width=\textwidth}
    \begin{tabular}{c|c|c|c|c|c|c|c|c|c|c|c}
    \hline
        \textbf{DATASET} & \makecell{\textbf{ALL} \\ \textbf{BANDS}} & \textbf{MVPCA \cite{mvpca}} & \textbf{FNGBS \cite{fn}} & \textbf{OCF \cite{ocf}} & \textbf{SR \cite{rank3}} & \textbf{SSR \cite{ssr}} & \textbf{GCSR \cite{graph}} & \textbf{SCDBS \cite{sparse}} & \makecell{\textbf{ABC-MI} \\ \textbf{(0.00)}} & \makecell{\textbf{ABC-MI} \\ \textbf{(0.01)}} & \makecell{\textbf{ABC-MI} \\ \textbf{(0.05)}} \\ \hline
        \textbf{SA} & 0.931±0.000 & 0.862±0.047 & 0.922±0.015 & 0.924±0.008 & 0.911±0.020 & 0.892±0.009 & 0.923±0.010 & 0.925±0.010 & 0.910±0.028 & 0.911±0.029 & 0.914±0.021 \\ \hline
        \makecell{\textbf{PA}} & 0.927±0.000 & 0.816±0.120 & 0.885±0.059 & 0.889±0.043 & 0.863±0.057 & 0.878±0.042 & 0.893±0.047 & 0.905±0.031 & 0.887±0.056 & 0.900±0.047 & 0.886±0.047 \\ \hline
    \end{tabular}
    \end{adjustbox}
    \label{tab2}
\end{table*}

Table \ref{tab1} and Table \ref{tab2} shows the average OA and Kappa values for different numbers of bands selected $(5:5:50)$ using our proposed method and other comparative methods using the SVM classifier on two datasets - PA and SA. The $(5:5:50)$ indicates the range of bands selected $(5-50)$ with an interval of $5$. Note that the values for MVPCA, FNGBS, OCF, SR, SSR, GCSR, and SCDBS are reported from the paper \cite{sparse}.

From the Table \ref{tab1} and Table \ref{tab2}, we can observe that the proposed ABC-MI method demonstrates competitive performance across both datasets, with its three variants showing distinct patterns. In PA, ABC-MI (0.01) outperforms the other variants (0.00 and 0.005) as well as other state-of-the-art methods and is only outperformed by the SCDBS method by a margin of 0.24$\%$. Compared to the case where all bands are used (ALL BANDS), the ABC-MI (0.01) method shows a decrease in overall accuracy (OA) of only 2.03\%, which lies well within its standard deviation of 3.36\%. This small difference demonstrates the effectiveness of the proposed approach in reducing spectral redundancy while maintaining accuracy comparable to using all available bands.

In terms of Kappa values, the ABC-MI (0.01) outperforms every other method except SCBDS, with a margin of only 0.005 and with ALL BANDS by a margin of 0.027, which is within its standard deviation value (0.047). In the Salinas dataset, the ABC-MI (0.05) outperforms SSR, SR, and MVPCA methods, and compared to the ALL BANDS OA value, it is outperformed by 0.95$\%$, which is within its standard deviation of 1.76$\%$. The Kappa value is also marginally lower by 0.017, again within its standard deviation. This demonstrates the effectiveness of the proposed ABC-MI method in terms of both OA and Kappa values. 

Table \ref{pcacomp} presents the overall accuracy of the proposed approach (best method) with that of PCA on the four datasets. It can be seen that a significantly higher accuracy is obtained for PA (from 86.25\% to 92.38\%) and SA (90.50\% to 92.23\%) datasets. For the LK dataset, a modest improvement is observed, while for the OS dataset, the performance remains comparable to that of PCA.

\begin{table*}[htbp]
    \centering
    \caption{\\COMPARISON IN TERMS OF OVERALL ACCURACY OF ABC-MI ($y$) WITH PCA}
    \renewcommand{\arraystretch}{2}
    \begin{tabular}{c|c|c}
    \hline
        \textbf{DATASET} & \textbf{ABC-MI BEST METHOD} & \textbf{PCA} \\ \hline
        \textbf{PA} & 92.38 ± 3.36 & 86.25 \\ \hline
        \textbf{SA} & 92.23 ± 1.76 & 90.50 \\ \hline
        \textbf{LK} & 96.68 ± 0.95 & 96.05 \\ \hline
        \textbf{OS} & 97.67 ± 0.08 & 97.72 \\ \hline
    \end{tabular}
    \label{pcacomp}
\end{table*}
\section{Conclusion}

Band selection is a crucial step in hyperspectral image analysis, as it involves extracting a smaller subset of bands to reduce spectral redundancy. Existing band selection approaches can be classified into three categories: ranking-based methods, clustering-based methods, and iterative or search-based methods. However, these methods are often prone to initialization problems, which can lead to sub-optimal band selections. They can also be sensitive to various parameter and hyper-parameter settings, requiring extensive fine-tuning and being computationally expensive.

 The present work explores a statistical dependence metrics-based approach to select the smaller subset of bands. In this work, the Variance Inflation Factor (VIF) is first utilized to pre-select bands with a focus on reducing multicollinearity and also the search space. Subsequently, Average Band Correlation (ABC) and Mutual Information (MI) are computed to quantify linear correlations between spectral bands, and also measure uncertainty reduction relative to ground truth labels. Finally, a clustering algorithm is used to identify the
optimal subset of bands based on the ABC and MI values. Unlike existing approaches, our method does not rely on any specific parameters, thus eliminating the need for optimal parameter estimation.

 The proposed approach is evaluated on four existing datasets, viz. WHU-
Hi-LongKou (LK), Oil Spill (OS), Pavia University (PA), and Salinas (SA) datasets, and compared with existing band selection approaches. The significant overlap of bands selected using our approach as compared to the existing approach demonstrates the effectiveness of our approach in reducing redundancy among bands. The ablation studies show that VIF-based on pruning effectively reduces multicollinearity. For instance, a reduction of 81\% to 45\% in the bands selected after VIF pre-screening on the LK dataset is observed depending on the VIF tolerance factor.  Indeed, the VIF tolerance factor incorporates a degree of tolerance in our work and introduces controlled collinearity in the pre-selection subset. The quality of the bands obtained using our method was assessed by using the support vector machine (SVM) classifier.  The ablation studies on the four datasets show that combining ABC with Mutual Information with VIF pre-screening yields robust and discriminative band subsets. 
In the Salinas dataset, the proposed approach achieved an overall accuracy of 92.23\% using the selected bands. In comparison, an overall accuracy of 93.18\% was obtained when all bands were considered. Furthermore, the 92.23\% accuracy on the Salinas dataset is comparable to existing state-of-the-art (SOTA) methods, such as SCDBS, which reported an accuracy of 93.15\%. A similar trend is observed in the Pavia dataset, where the proposed method yielded an overall accuracy of 92.38\% with the selected bands, compared to 94.41\% when using all bands and 92.62\% for the SCDBS method. These results demonstrate the effectiveness of the proposed approach in reducing spectral redundancy while mitigating multicollinearity. 

\small
\bibliographystyle{IEEEtranN}
\bibliography{ref}

\end{document}